\documentclass[pra,twocolumn,twoside]{revtex4}

\usepackage{graphicx}
\usepackage{amsmath,amssymb}
\usepackage{ae}

\newcommand{\eq}[1]{\begin{equation}#1\end{equation}}
\newcommand{\eqmulti}[1]{\begin{equation}\begin{split}#1\end{split}\end{equation}}

\newcommand{\ket}[1]{\ensuremath{\,|{#1}\rangle}}

\newcommand{\matrixe}[3]{\ensuremath{\langle{#1}|\,{#2}\,|{#3}\rangle}}

\newcommand{\op}[1]{\ensuremath{\hat{\mathrm{#1}}}}
\newcommand{\adj}[1]{\ensuremath{{{#1}}^{\dag}}}

\newcommand{\aO}{\ensuremath{\op{a}}}
\newcommand{\aaO}{\ensuremath{\adj{\op{a}}}}
\newcommand{\cO}{\ensuremath{\op{c}}}
\newcommand{\ccO}{\ensuremath{\adj{\op{c}}}}
\newcommand{\nO}{\ensuremath{\op{n}}}

\newcommand{\HO}{\ensuremath{\op{H}}}

\newcommand{\vV}{\ensuremath{\vec{v}}}
\newcommand{\xV}{\ensuremath{\vec{x}}}

\newcommand{\nablaV}{\ensuremath{\vec{\nabla}}}

\newcommand{\ii}{\ensuremath{\mathrm{i}}}
\newcommand{\ee}{\ensuremath{\mathrm{e}}}

\newcommand{\SF}{\ensuremath{\textrm{s}}}

\newcommand{\CN}{\ensuremath{\textrm{d}}}

\newcommand{\B}{\ensuremath{\textrm{B}}}
\newcommand{\F}{\ensuremath{\textrm{F}}}
\newcommand{\BB}{\ensuremath{\textrm{BB}}}
\newcommand{\BF}{\ensuremath{\textrm{BF}}}

\begin{document}

\title{Quantum phases of atomic boson-fermion mixtures in optical lattices}

\author{Robert Roth}
\affiliation{Institut f\"ur Kernphysik, Technische Universit\"at Darmstadt,
64289 Darmstadt, Germany}

\author{Keith Burnett}
\affiliation{Clarendon Laboratory, Department of Physics, 
University of Oxford, Parks Road, Oxford OX1 3PU, United Kingdom}

\date{\today}

\begin{abstract}    
The zero-temperature phase diagram of a binary mixture of bosonic and
fermionic atoms in an one-dimensional optical lattice is studied in the
framework of the Bose-Fermi-Hubbard model. By exact numerical solution of
the associated eigenvalue problems, ground state observables and the
response to an external phase twist are evaluated. The stiffnesses under
phase variations provide measures for the boson superfluid fraction and
the fermionic Drude weight. Several distinct quantum phases are identified
as function of the strength of the repulsive boson-boson and the
boson-fermion interaction. Besides the bosonic Mott-insulator phase, two
other insulating phases are found, where both the bosonic superfluid
fraction and the fermionic Drude weight vanish simultaneously. One of
these double-insulator phases exhibits a crystalline diagonal long-range
order, while the other is characterized by spatial separation of the two
species. 
\end{abstract}



\maketitle


Following the seminal experiments on the superfluid to Mott-insulator
transition in ultracold atomic Bose gases in optical lattice potentials
\cite{GrMa02a,GrMa02b}, a new frontier in the field of degenerate, strongly
correlated quantum gases has emerged. It has recently become possible to
prepare degenerate mixtures of a bosonic and a fermionic atomic species in
an optical lattice \cite{MoFe03}. These degenerate gases will facilitate the
experimental study of quantum phase transitions in systems of mixed quantum
statistics, which are extremely hard to access in the solid-state context.
The superior experimental control available in ultracold atomic gases
experiments promises detailed insights into the rich physics of quantum 
phase transitions. In this paper we explore the phase diagram of a binary
boson-fermion mixture in an optical lattice and identify experimental
signatures of the different phases.


A theoretical framework for degenerate boson-fermion mixtures in optical
lattices can be constructed by a straight-forward generalization of the
single-band Bose-Hubbard model, used successfully to describe the
superfluid to Mott-insulator phase transition in purely bosonic systems
\cite{JaBr98}. A complete single-particle basis within the lowest Bloch band 
is given by the Wannier functions for the vibrational ground state of each
lattice well. We define creation operators $\aaO_i$ ($\ccO_i$) which
create a boson (fermion) in the localized Wannier state at site
$i=1,...,I$. Using this basis we can translate the many-body Hamiltonian
into a second quantized form. This leads to the Bose-Fermi-Hubbard (BFH)
Hamiltonian for a one-dimensional translationally invariant lattice
\cite{AlIl03,BuBl03,LeSa03}
\eqmulti{ \label{eq:hamiltonian}
  \HO
  =& -J_{\B} \sum_{i=1}^{I} (\aaO_{i+1} \aO_{i} + \textrm{h.a.})
  - J_{\F} \sum_{i=1}^{I} (\ccO_{i+1} \cO_{i} + \textrm{h.a.}) \\
  &+ \frac{V_{\BB}}{2} \sum_{i=1}^{I} \nO^{\B}_i (\nO^{\B}_i-1)
  + V_{\BF} \sum_{i=1}^{I} \nO^{\B}_i \nO^{\F}_i \;,
}
where $\nO^{\B}_i= \aaO_i \aO_i$ and $\nO^{\F}_i= \ccO_i \cO_i$ are the
occupation number operators for bosons and fermions, respectively. The
first line describes the tunneling between adjacent lattice sites. The
tunneling amplitudes $J_{\B}$ and $J_{\F}$ are connected to off-diagonal
matrix elements of the kinetic energy and the lattice potential
\cite{AlIl03}. The second line of Eq. \eqref{eq:hamiltonian} contains the
on-site boson-boson and boson-fermion interactions with interaction
strengths $V_{\BB}$ and $V_{\BF}$, respectively. Within the single-band
approximation a fermion-fermion on-site interaction does not arise as the
Pauli principle prevents two identical fermions from occupying the same
site. 

We exactly solve the eigenvalue problem of the BFH Hamiltonian within a
complete basis of Fock states of the form
$\ket{\{n^{\B}_1,...,n^{\B}_I\}_{\alpha}}\otimes\ket{\{n^{\F}_1,...,n^{\F}_I\}_{\beta}}$.
By $\{n^{\B}_1,...,n^{\B}_I\}_{\alpha}$ we denote a set of boson
occupation numbers for the individual sites. The index
$\alpha=1,...,D_{\B}$ labels all possible compositions of occupation
numbers which yield $\sum_{i=1}^{I} n^{\B}_i = N_{\B}$. For the set   of
fermion occupation numbers $\{n^{\F}_1,...,n^{\F}_I \}_{\beta}$ only those
compositions with $0$ or $1$ fermions at a given site are allowed by the
Pauli principle. The dimensions of the boson and fermion space are given
by $D_{\B} = (N_{\B}+I-1)!/(N_{\B}! (I-1)!)$ and $D_{\F} =  I!/(N_{\F}!
(I-N_{\F})!)$, respectively. For a lattice with $I=8$ sites, $N_{\B}=8$
bosons, and $N_{\B}=4$ fermions the basis dimensions are $D_{\B}=6435$ and
$D_{\F}=70$. Clearly, the dimension of the fermion space is dramatically
reduced due to the Pauli principle.

Within the basis of Fock-states we construct the matrix representation of
the BFH Hamiltonian assuming periodic boundary conditions. The lowest
eigenvalues and eigenvectors of the Hamilton matrix are computed using a
refined Lanczos algorithm. The eigenvectors provide the coefficients for
the representation of the eigenstates $\ket{\Psi_{\nu}}$ in the Fock-state
basis
\eq{ \label{eq:state}
  \ket{\Psi_{\nu}} 
  = \sum_{\alpha=1}^{D_{\B}} \sum_{\beta=1}^{D_{\F}}  
    C^{(\nu)}_{\alpha\beta} \ket{ \{n^{\B}_1,...,n^{\B}_I\}_{\alpha} }
    \otimes \ket{\{n^{\F}_1,...,n^{\F}_I \}_{\beta} } \;.
}
From this representation of the ground state we can extract various
observables such as mean occupation numbers and number fluctuations
for both species. It also provides access to more elaborate
quantities such as the one- or two-body density matrix and the static
structure factor, which we will discuss below.

The dynamical response of the system to external perturbations is a 
crucial quantity in distinguishing between different quantum phases, for
example, between the bosonic Mott-insulator and the superfluid phase. By
imposing a linear variation of the phase onto the many-body wavefunction
one can probe the mobility of the atoms in the lattice. On the single
particle level, a spatial variation of the phase of the wavefunction is
associated with a velocity field $\vV(\xV) = \frac{\hbar}{m}
\nablaV\theta(\xV)$, and a linear phase variation thus corresponds to a
constant velocity across the lattice. Roughly speaking, those particles
free to move will respond to the phase twist and exhibit a homogeneous
flow. This flow is then associated with an additional kinetic energy and
hence an increase of the total energy. This provides a measure for the
density of mobile particles in the lattice \cite{RoBu03c,RoBu03a}.

This phase variation can be realized by imposing twisted boundary
conditions for the many-body wavefunction. In the case of the BFH model it
is, however, more convenient to map the phase twist by means of a unitary
transformation onto the Hamiltonian \cite{RoBu03c}. This leads to a
``twisted'' Hamiltonian which contains  Peierls phase factors in the
hopping terms: 
\eqmulti{ \label{eq:hamiltonian_twist}
  \HO_{\Theta_{\B},\Theta_{\F}}
  =& -J_{\B} \sum_{i=1}^{I} (\ee^{-\ii \Theta_{\B}/I}\;
    \aaO_{i+1} \aO_{i} + \textrm{h.a.}) \\
  &- J_{\F} \sum_{i=1}^{I} (\ee^{-\ii \Theta_{\F}/I}\;
    \ccO_{i+1} \cO_{i} + \textrm{h.a.}) \\
  &+ \frac{V_{\BB}}{2} \sum_{i=1}^{I} \nO^{\B}_i (\nO^{\B}_i-1)
    + V_{\BF} \sum_{i=1}^{I} \nO^{\B}_i \nO^{\F}_i \;.
}
In order to probe the response of the bosonic and the fermionic species
separately, we have introduced different total twist angles $\Theta_{\B}$ and
$\Theta_{\F}$, respectively. 

For an isolated boson twist ($\Theta_{\B}>0, \Theta_{\F}=0$) the
difference in the ground state energies of the boson-twisted Hamiltonian
$\HO_{\Theta_{\B},0}$ and the initial Hamiltonian $\HO_{0,0}$ provides
direct information on the superfluid density of the bosonic component.
This is because the superfluid component responds to the imposed phase gradient
producing flow and the associated energy change. The bosonic phase stiffness 
\eq{
  f^{\B}_{\SF}
  = \frac{I^2}{N_{\B} J_{\B}}\; \frac{E_{\Theta_{\B},0} -
  E_{0,0}}{\Theta_{\B}^2} \;,\quad \Theta_{\B}\ll\pi 
}
is, therefore, a measure for the boson superfluid fraction \cite{RoBu03a, RoBu03c,
FiBa73, Krau91}. We note that neither the  stability of the superflow nor
the reduction of the flow by the lattice itself are taken into account by
this quantity. The superfluid weight $f^{\B}_{\SF}$ vanishes for
insulating phases such as the Mott-insulator phase and goes to $1$ for a
perfect superfluid.

The energy change resulting from an isolated fermion twist 
($\Theta_{\B}=0, \Theta_{\F}>0$) provides a measure for the mobility of
the fermionic atoms. The fermionic phase stiffness
\eq{
  f^{\F}_{\CN}
  = \frac{I^2}{N_{\F} J_{\F}}\; \frac{E_{0,\Theta_{\F}} -
  E_{0,0}}{\Theta_{\F}^2}  \;,\quad \Theta_{\F}\ll\pi
}
is equivalent to the well-known Drude weight for charged fermions in a
lattice \cite{FyMa91,ScWh93,ShSu90}. The vanishing of the Drude weight is
an indicator of an insulating state, whereas a finite value indicates
non-vanishing conductivity.


These two phase stiffnesses are an important tool to characterize the
different quantum phases of the boson-fermion mixture in a lattice. In the
following we map out the phase diagrams for different boson and fermion
numbers, $N_{\B}$ and $N_{\F}$, as function of the strength of the
repulsive boson-boson and the boson-fermion interaction. For simplicity we
examine the case of equal tunneling rates for the two species, i.e.
$J=J_{\B}=J_{\F}$, and use $J$ as energy unit. The only remaining
parameters of BFH Hamiltonian are then the ratios between the two
interaction strengths and the tunneling rate, $V_{\BB}/J$ and $V_{\BF}/J$.
For each point in the $V_{\BB}$-$V_{\BF}$-plane we solve the eigenvalue
problem of the BFH Hamiltonian \eqref{eq:hamiltonian} and the two twisted
Hamiltonians $\HO_{\Theta_{\B},0}$ and $\HO_{0,\Theta_{\F}}$ (with
$\Theta_{\B},\Theta_{\F}=0.1$) in order to compute the phase stiffnesses
and several simple ground state observables.


\begin{figure*}
\includegraphics[width=0.85\textwidth]{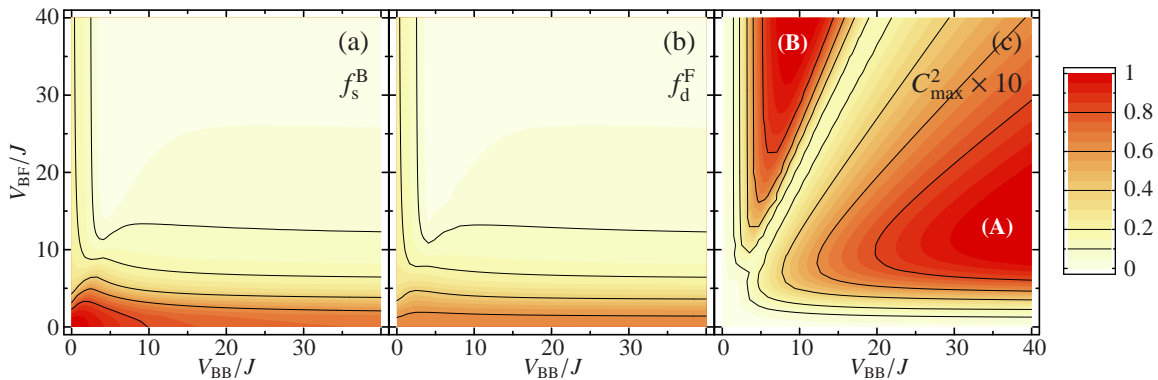}
\vspace*{-2ex}
\caption{Contour plots of (a) the bosonic superfluid fraction
$f_{\SF}^{\B}$, (b) the fermionic Drude weight $f_{\CN}^{\F}$, and (c) the
largest coefficient $C_{\max}^2$ as function of $V_{\BB}/J$ and $V_{\BF}/J$
for a lattice with $I=8$ sites and $N_{\B}=N_{\F}=4$.}
\label{fig:4+4_phasediag}
\end{figure*}

First we consider a lattice with $I=8$ sites and equal filling factors 
$N_{\B}/I=1/2$ and $N_{\F}/I=1/2$ for the bosonic and the fermionic
species, respectively. Although the system is rather small we could confirm
that finite size effects are insignificant for the observables discussed in the
following \cite{RoBu03c}. Figure \ref{fig:4+4_phasediag} shows contour
plots of the boson stiffness $f_{\SF}^{\B}$, i.e., the superfluid fraction
within the bosonic species, and the fermion stiffness $f_{\CN}^{\F}$,
i.e., the Drude weight for the fermionic component. In addition the
largest coefficient $C_{\max}^2$ of the expansion \eqref{eq:state} of the
ground state is given. Large values of $C_{\max}^2$ indicate that the ground
state is dominated by one or a few particular Fock states. 

Figures \ref{fig:4+4_phasediag}(a) and (b) reveal that in the large
parameter region for $V_{\BB}/J \gtrsim5$ and $V_{\BF}/J \gtrsim 15$ the
boson and the fermion phase stiffnesses vanish. The bosonic and the
fermionic components then behave like insulators, i.e., the superfluid
fraction and the Drude weight are both zero. This double-insulator phase
is generated by the combined action of the boson-boson and the
boson-fermion repulsion. Any configuration which has two bosons or a boson
and a fermion at the same site is energetically unfavorable, and the
ground state is therefore dominated by Fock states with either one boson or one
fermion per site. Only if one of the interaction strengths is small can a
finite stiffnesses for bosons and fermions emerge.

Closer inspection of the largest coefficient of the expansion
\eqref{eq:state}, displayed in Fig. \ref{fig:4+4_phasediag}(c), reveals that within the
double-insulating region there are two distinct areas where the largest
coefficient reaches $C_{\max}^2\approx0.1$. Hence there is a class of Fock
states which provide the dominant contribution to the ground state. In
region (A) for values $2 V_{\BB}\gtrsim V_{\BF}$ the dominant Fock states
exhibits an alternating occupation of the sites with one boson or one
fermion. In region (B) the dominant Fock states contain a continuous block
of bosons followed by a block of fermions. Note that although these states
have the largest weight, other Fock states do also contribute to the
ground state. Most important are those which can be generated from the
dominant Fock state by exchange of two lattice sites or particle-hole
excitations.

The reason for the existence of and the cross-over between these two
different intrinsic structures is quite subtle. All the interaction matrix
elements for the dominant Fock states are irrelevant, because all these states
have exactly one particle (boson or fermion) at each site. The existence
of the two different structures is triggered by states with double
occupancies that enter into the ground state with much smaller weight but
nevertheless couple to the dominant states via the tunneling term. This
complex correlation dictates the cross-over between alternating and
separated species.

\begin{figure}
\includegraphics[width=0.33\textwidth]{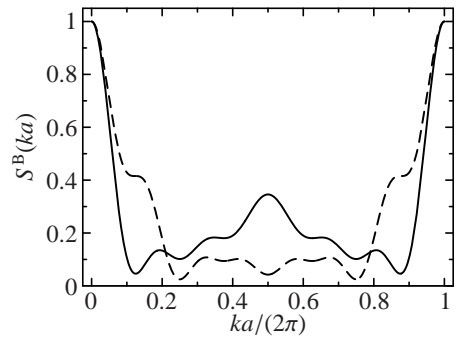}
\vspace*{-2ex}
\caption{Static structure factor $S^{\B}(ka)$ for $I=8$ and $N_{\B}=N_{\F}=4$ as function of $ka$ for the
ground state with alternating occupation at $V_{\BB}/J=40$,
$V_{\BF}/J=15$  (solid line) and with separated species for
$V_{\BB}/J=10$, $V_{\BF}/J=40$ (dashed line).}
\label{fig:4+4_strucfac}
\end{figure}

A good observable to distinguish the different intrinsic configurations
is the static structure factor \cite{RoBu03c}. For the bosonic component
it is given by
\eq{
  S^{\B}(k a) 
  = \frac{1}{N_{\B}^2} \sum_{i,j=1}^{I} \ee^{\ii k a (i-j)}
    \matrixe{\Psi_0}{\nO^{\B}_i \nO^{\B}_j}{\Psi_0} \;,
}
where $a$ is the lattice spacing. It measures the density-density
correlations in the system and thus provides information on the presence
or otherwise of diagonal long-range order. 

Figure \ref{fig:4+4_strucfac} shows an example for the structure factor
within the region of alternating occupation (full line) and the region of
phase separation (dashed line). Apart from the trivial peaks at integer
multiples of $2\pi$ the structure factor differs substantially within the
two phases. For the alternating occupation phase, $S^{\B}(k a)$ exhibits
an enhancement around $ka=\pi$ which signals an increased probability for
a boson to be found at every other site. This is the signature of diagonal
long-range order as it is found in a crystal. In a two-dimensional lattice
the corresponding checkerboard configuration, which exhibits crystalline
diagonal long-range order, was identified recently
\cite{BuBl03}. 

Within the separated phase the structure factor shows a rather different
behavior (dashed line in Fig. \ref{fig:4+4_strucfac}): it is suppressed
around $ka=\pi$ and enhanced at $ka \approx 2\pi/I$. Due to this
difference in the structure factors it might be possible to distinguish
the two types of double-insulator phases experimentally, e.g., through
selective Bragg diffraction of light off the trapped atoms. We note that
the interference pattern produced by the atoms after release from the
lattice, along with ballistic expansion, will not distinguish the phases as
their coherence properties, i.e., the off-diagonal long-range order, are
too similar.


\begin{figure*}
\includegraphics[width=0.85\textwidth]{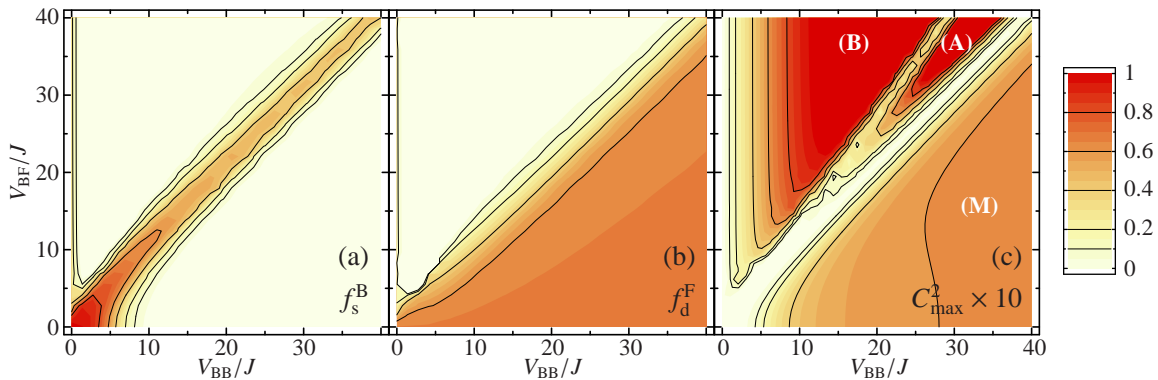}
\vspace*{-2ex}
\caption{Contour plots of (a) the bosonic superfluid fraction
$f_{\SF}^{\B}$, (b) the fermionic Drude weight $f_{\CN}^{\F}$, and (c) the
largest coefficient $C_{\max}^2$ as function of $V_{\BB}/J$ and $V_{\BF}/J$
for a lattice with $I=8$ sites, $N_{\B}=8$, and $N_{\F}=4$.}
\label{fig:8+4_phasediag}
\end{figure*}

\enlargethispage{3ex}

As a second example we consider a lattice with $I=8$ sites, a boson
filling factor $N_{\B}/I=1$, and a fermion filling factor $N_{\F}/I=1/2$.
Contour plots of the bosonic and fermionic stiffness and the largest
coefficient as function of $V_{\BB}/J$ and $V_{\BF}/J$ are shown in Fig.
\ref{fig:8+4_phasediag}.  

The qualitative difference compared to the half-filling case is the
appearance of a bosonic Mott-insulator phase (M) for $V_{\BB}>V_{\BF}$. In
this region the boson stiffness, i.e. the superfluid fraction, vanishes
whereas the fermionic stiffness remains at its noninteracting value as
depicted in Figs. \ref{fig:8+4_phasediag}(a) and (b). Under the influence
of the strong boson-boson repulsion the bosons form a Mott insulating
layer. The fermions are not affected, they experience a homogeneous
background of bosons on which they can move freely.

The situation changes if the boson-fermion interaction becomes comparable
in strength to the boson-boson interaction. Through the boson-fermion
repulsion the fermions are able to break up the bosonic Mott-insulator and
partially restore bosonic superfluidity in the region $V_{\BF}\approx
V_{\BB}$. Further increase of the boson-fermion repulsion leads to the
simultaneous reduction of both phase stiffnesses and to the formation of a
double-insulator regime as observed in the case of half-filling. The
contour plot for the largest coefficient $C_{\max}^2$ depicted in Fig.
\ref{fig:8+4_phasediag}(c) reveals that both of the previously observed 
intrinsic configurations appear again: (A) an alternating occupation of 
the sites associated with crystalline diagonal long-range order and (B)
continuous blocks of bosons and fermions which resemble a spatial
separation of the two components. Due to the appearance of the bosonic
Mott-insulator both phases have been shifted to the region
$V_{\BF}>V_{\BB}$. 


In summary, we have mapped out the phase diagrams of a binary mixture of
bosonic and fermionic atoms in a one-dimensional lattice. By exact
numerical solution of the eigenvalue problem for the Bose-Fermi-Hubbard
Hamiltonian we are able to compute ground states as well as the dynamical
response of the system. The stiffnesses of the bosonic and fermionic
component under phase twists provides an important measure for the
superfluid properties of the bosonic component (superfluid fraction) and
the conductivity of the fermionic species (Drude weight). Using these
quantities we have identified several distinct quantum phases, e.g., a
bosonic Mott-insulator phase and two double-insulator phases which exhibit
different degrees of diagonal long-range order---alternating occupation
and component separation. This shows that atomic boson-fermion mixtures in
optical lattices constitute a versatile tool for the study of fundamental
quantum phase transitions. 


This work was supported by the DFG, the UK EPSRC, and the EU via the
``Cold Quantum Gases'' network. K.B. thanks the Royal Society and
Wolfson Foundation for support.


\end{document}